\UseRawInputEncoding
\documentclass[superscriptaddress, twocolumn, amsmath, amssymb, aps,pra, notitlepage,longbibliography]{revtex4-2} 
\usepackage{graphicx,graphics,epsfig,subfigure,times,bm,bbm,amssymb,amsmath,amsfonts,amsthm,mathrsfs,MnSymbol}
\usepackage[matrix,frame,arrow]{xypic}
\usepackage[normalem]{ulem}
\usepackage{slashed}
\usepackage{dcolumn}
\usepackage{tabularx}
\usepackage{amsopn}
\usepackage{color}
\usepackage[usenames,dvipsnames,svgnames,table]{xcolor}
\usepackage[english]{babel}
\usepackage{verbatim}
\definecolor{darkblue}{rgb}{0.0,0.0,0.3}
\usepackage[colorlinks=true,
            linkcolor=red,
            urlcolor= darkblue,
            citecolor=blue]{hyperref}

\newcommand{\bea}{\begin{eqnarray}}
\newcommand{\eea}{\end{eqnarray}}

\begin{document}

\title{Dynamical non-Hermitian systems: Fingerprints and pure dephasing induced protection effect}

\author{Wanchen Ma}
\affiliation{Department of Physics, Institute for Quantum Science and Technology, International Center of Quantum and Molecular Structures, Shanghai University, Shanghai, 200444, China}
\author{Hao Zhang}
\email{zhangh@fudan.edu.cn}
\affiliation{School of Information Science and Technology, Key Laboratory for Information Science of Electromagnetic Waves (MOE), Department of Optical Science and Engineering, Key Laboratory of Micro and Nano Photonic Structures (MOE), Fudan University, Shanghai 200433, China}
\affiliation{Yiwu Research Institute of Fudan University,
Chengbei Road, Yiwu City, Zhejiang 322000, China}
\author{Junjie Liu}
\email{jj\_liu@shu.edu.cn}
\affiliation{Department of Physics, Institute for Quantum Science and Technology, International Center of Quantum and Molecular Structures, Shanghai University, Shanghai, 200444, China}

\begin{abstract}
Non-Hermitian systems have shown promising potential for realizing quantum information tasks that lack counterparts in the Hermitian realm. Understanding the dynamical characteristics of non-Hermitian systems as reflected in information-theoretic quantities is essential for advancing their applications. Here we investigate dynamics of trace distance and concurrence, which quantify information flow and entanglement, respectively, in non-Hermitian qubit systems exhibiting either parity-time or anti-parity-time symmetry. We identify dynamical fingerprints from the system density matrix which provide a unified explanation to the common oscillation and relaxation dynamics shared by trace distance and concurrence observed in existing studies. We further investigate the fate of dynamical fingerprints under the impact of pure dephasing. Surprisingly, we find that pure dephasing can slow down the inherent relaxation of information flow and entanglement in non-Hermitian systems with unbroken anti-parity-time symmetry, suggesting a passive environment-assisted information protection in this class of non-Hermitian systems. Our findings enhance our understanding of dynamical non-Hermitian systems and have specific relevance to their information-oriented applications.
\end{abstract}

\date{\today}
\maketitle


\section{Introduction}\label{sec:1}
Non-Hermitian systems have recently attracted considerable theoretical and experimental interest  \cite{bender.2007.rpp,Ganainy.18.NP,Ashida.20.AP,Bergholtz.21.RMP,Okuma.23.ARCMP}. Depending on the specific symmetry involved, two primary classes of non-Hermitian systems are currently under active investigation: those exhibiting parity-time (PT) symmetry \cite{Bender.1998.prl,bender.2005.cp,Bender.24.RMP} and those supporting anti-parity-time (APT) symmetry \cite{peng.2016.np,choi.2018.nc,park.2021.prl,Lic.24.PRL}. With the advent of these now highly adopted concepts, a plethora of novel phenomena that have no counterparts in Hermitian settings have been unraveled. Notable examples include the non-Hermitian skin effect \cite{Yao.18.PRL,Yokomizo.19.PRL,Lee.19.PRB,Longhi.19.PRR,Zhang.20.PRL,Okuma.20.PRL,Lin.23.FP},  non-Hermitian topology \cite{Rudner.09.PRL,Esaki.11.PRB,Malzard.15.PRL,Lee.16.PRL,Leykam.17.PRL,Kunst.18.PRL,Xiong.2018.JPC,Yao.18.PRl2,Song.19.PRl,Imura.19.PRL,Edvardsson.19.PRL,Gong.18.PRX}, exceptional points \cite{mandal.2021.prl,miri.2019.science,Ozdemir.19.NM}, chiral state transfer \cite{Zhang.12.PRA,ren.2022.np,SunK.23.PRA,Arkhipov.24.PRL} and enhanced non-Hermitian sensing \cite{fleury.2015.nc,sakhdari.2019.prl,Tchodimou.2017.pra,liu.2016.prl,li.2016.pra}, to name just a few.

These intriguing features indicate that dynamical non-Hermitian systems hold significant potential for finite-time quantum information processing applications. Therefore, a comprehensive understanding of dynamical behaviors of non-Hermitian systems in relation to information-theoretic quantities becomes essential. To date, various studies have reported dynamical aspects related to information-theoretic quantities in non-Hermitian systems. Notably, the dynamics of entanglement between non-Hermitian qubits can exhibit periodic oscillation or exponential decay depending on whether the symmetry is broken \cite{Nori.2022.prr,akram.2023.sr}. Interestingly, the dynamical evolution of information flow \cite{kawabata.2017.prl,P.Xue.2019.prl,Wen.20.NPJQI} and information content \cite{DingYL.2022.pra} also exhibits similar dynamical trends. Given the distinct nature of the quantities addressed, one naturally wonders whether there are common dynamical fingerprints underpinning the observed similar dynamical behaviors. However, existing studies are often confined to dynamical behaviors of specific information-theoretic quantities in non-Hermitian systems with a specified symmetry, making a general understanding of the common dynamical behaviors still lacking.  

Furthermore, it should be noted that realistic non-Hermitian systems are typically fabricated by truncating larger quantum systems \cite{Ozdemir.19.NM,chenwj.2021.prl,chenwj.2022.prl,ren.2022.np,sun.2023.pra}. The additional degrees of freedom of the larger systems, usually in the form of extra energy levels, not only induce gain and loss needed to realize the desire non-Hermiticity but also introduce additional dissipative effects to the non-Hermitian systems. Given this context, realistic non-Hermitian systems should be treated as open ones. In the presence of dissipation, one typically expects a fragility in information-theoretic quantities, as additional dissipation can degrade their quantum features \cite{breuer.2002.book} and impact their dynamical behaviors. Therefore, understanding how dynamical non-Hermitian systems respond to additional dissipative effects, such as pure dephasing \cite{Cummings.16.PRl,Chesi.16.PRL,Cen.22.PRA}, is equally important for developing their information-oriented applications.

To address these issues, here we aim to achieve two objectives: (i) identifying the dynamical fingerprints underlying the common dynamical behaviors shared by distinct information-theoretic quantities in closed non-Hermitian systems, and (ii) examining the dynamical response of these quantities to additional pure dephasing effects. To achieve the first objective, we note that previously addressed information-theoretic quantities \cite{Nori.2022.prr,akram.2023.sr,kawabata.2017.prl,P.Xue.2019.prl,DingYL.2022.pra} build solely upon the system density matrix. Thus, we focus on the dynamical evolution of the system density matrix {which we show} can either be periodic or experience exponential decay, with analytical expressions for the oscillation period [cf. Eq. (\ref{closed_T})] and relaxation time [cf. Eq. (\ref{closed_tau})] obtained. We numerically confirm the validity of these analytical expressions in describing the similar dynamical behaviors shared by the trace distance \cite{kawabata.2017.prl,P.Xue.2019.prl,nielsen.2001.book} and the concurrence \cite{Nori.2022.prr,akram.2023.sr,wootters.1998.prl} using two-qubit non-Hermitian systems with PT or APT symmetry, in  both the symmetry-unbroken and -broken phases. Thus, we identify the behaviors governed by Eqs. (\ref{closed_T}) and (\ref{closed_tau}) as general dynamical fingerprints of non-Hermitian systems, regardless of the symmetry type, which provide a unified understanding of existing studies \cite{Nori.2022.prr,akram.2023.sr,kawabata.2017.prl,P.Xue.2019.prl,DingYL.2022.pra}.

To incorporate the pure dephasing effect, we further utilize a quantum master equation approach \cite{breuer.2002.book} to describe the evolution of open non-Hermitian systems. Using two-qubit non-Hermitian systems as numerical examples, we find that the pure dephasing--whether local or collective--induces detrimental effects in the parameter regime where information-theoretic quantities depict dynamical oscillation in closed settings, irrespective of the symmetry type, similar to the scenario of Hermitian systems under pure dephasing. Whereas in the dynamical regime where the dynamics of information-theoretic quantities in closed setting already follow decaying trends, we observe that non-Hermitian systems with different symmetries would have distinct responses to the pure dephasing. Specifically, in non-Hermitian systems with broken PT symmetry, pure dephasing accelerates the relaxation processes of information-theoretic quantities. In stark contrast, we reveal a surprising slow-down effect of pure dephasing in non-Hermitian systems with unbroken APT symmetry. Here, the relaxation time can increase with the strength of pure dephasing, suggesting a beneficial role of the typically detrimental pure dephasing in protecting the corresponding non-Hermitian systems from losing quantum features. 

The structure of the paper is as follows. In Sec. \ref{sec:2}, we first examine the dynamical evolution of the normalized density matrix for closed non-Hermitian systems. We identify different dynamical regimes based on whether the energy eigenvalues are pure real or complex, and we obtain analytical expressions for numerical benchmarks. We then assess the utility of these analytical insights by comparing them with the simulated dynamics of the trace distance and the concurrence, which quantify the information flow and entanglement in the non-Hermitian systems, respectively. These comparisons ultimately lead to a unified understanding of existing results by identifying underlying dynamical fingerprints. In Sec. \ref{sec:3}, we shift focus to open non-Hermitian systems and examine the pure dephasing effect. We derive a quantum Lindblad-type master equation that describes the evolution of the normalized density matrix for open non-Hermitian systems. We numerically analyze the dynamical behaviors of the trace distance and the concurrence in two-qubit non-Hermitian systems with either unbroken or broken symmetries under varying strength of pure dephasing, particularly focusing on an unexpected slow-down effect induced by pure dephasing and discuss its extent. We conclude the study in Sec. \ref{sec:4}.

\section{Dynamical fingerprints of closed non-Hermitian systems} \label{sec:2}
We begin with dynamical behaviors of closed non-Hermitian systems upon which existing studies on information-theoretic quantities are based. To facilitate comparisons with existing results, we utilize the time-dependent normalized density matrix to describe the time evolution of closed non-Hermitian systems \cite{brody2012mixed,cao.2023.prr,du.2022.pra,P.Xue.2019.prl,Nori.2022.prr,Yuan.20.PRL,Mao.24.CPL}
\begin{equation}
\hat{\rho}(t)=\frac{e^{-i\hat{H}t}\hat{\rho}(0)e^{i\hat{H}^\dagger t}}{\mathrm{Tr}[e^{-i\hat{H}t}\hat{\rho}(0)e^{i\hat{H}^\dagger t}]}.
\label{closed_H}
\end{equation}
Here, $\hat{H}$ denotes the Hamiltonian of the closed non-Hermitian system with $\hat{H}^{\dagger}$ representing its complex conjugate. From this form, the equation of motion for the normalized density matrix can be readily expressed as \cite{brody2012mixed}
\begin{equation}
\frac{d}{dt}\hat{\rho}(t)=-i\left[\hat{H}\hat{\rho}(t)-\hat{\rho}(t)\hat{H}^{\dagger} \right]-i\mathrm{Tr}[\hat{\rho}(t)(\hat{H}^{\dagger}-\hat{H})]\hat{\rho}(t).
\label{closed_L}
\end{equation}
The second term on the right-hand-side of Eq. (\ref{closed_L}) represents a correction term that ensures the conservation of probability $\mathrm{Tr}[\hat{\rho}(t)]=1$ in non-Hermitian systems. When $H=H^{\dagger}$, Eq. (\ref{closed_L}) reduces to the conventional Liouville equation for closed Hermitian systems as expected.

We observe that the dynamics of information-theoretic quantities examined in previous studies \cite{Nori.2022.prr,akram.2023.sr,kawabata.2017.prl,P.Xue.2019.prl,DingYL.2022.pra} are solely induced by the evolution of system density matrix, since measuring the information encoded within the system state is of primary interest. Therefore, to extract the underlying dynamical fingerprints behind the similar dynamical behaviors shared by those information-theoretic quantities, we focus on the dynamical evolution of the density matrix in Eq. (\ref{closed_H}). We employ a spectral decomposition approach \cite{kawabata.2017.prl} which enables us to decompose the density matrix in Eq. \eqref{closed_H} as
\begin{equation}
\hat{\rho}\left(t\right)=\frac{\sum\limits_{mn}\rho_{mn}e^{-i\omega_{mn}t}\left|\varphi_{m}\right\rangle\left\langle\varphi_{n}\right|}{\mathrm{Tr}\left[\sum\limits_{mn}\rho_{mn}e^{-i\omega_{mn}t}\left|\varphi_{m}\right\rangle\left\langle\varphi_{n}\right|\right]}.
\label{rhosd}
\end{equation}
Here, $\{\left|\varphi_{m}\right\rangle\}$ ($\{\left\langle \chi_{m}\right|\}$) are the right (left) eigenstates of $\hat{H}$ with corresponding eigenvalues $\lambda_m=E_m+i\Gamma_m$, we have denoted $\omega_{mn}\equiv\lambda_{m}-\lambda^{\ast}_{n}$ with $\lambda^{\ast}_{n}$ being the complex conjugate of $\lambda_n$ and $\rho_{mn}=\frac{\langle\chi_m|\hat{\rho}_0|\chi_n\rangle}{\langle\chi_m|\varphi_m\rangle\langle\varphi_n|\chi_n\rangle}$ with $\hat{\rho}_0$ the initial state. For convenience, we arrange eigenvalues $\{\lambda_m\}$ in descending order of $\Gamma_m$, such that $\Gamma_{1} \geq \Gamma_{2} \geq \Gamma_{3} \geq \cdots$.

In the long-time limit, Eq. \eqref{rhosd} simplifies to (see details in Appendix \ref{a:1})
\begin{equation}
\begin{gathered}
    \hat{\rho}(t)~\simeq~|\varphi_{1}\rangle\langle\varphi_{1}|  +\Theta(t) e^{-\Delta\Gamma t}.
    \label{closed_rho}
\end{gathered}
\end{equation}
Here, we have defined $\Theta(t)=\left(R_1e^{-i\Delta E t}|\varphi_{1}\rangle\langle\varphi_{2}|+\mathrm{h.c.}\right)  - \left(R_1e^{-i\Delta E t}\langle\varphi_{2}|\varphi_{1}\rangle+\mathrm{c.c.}\right)|\varphi_{1}\rangle\langle\varphi_{1}|$ with $R_1=\rho_{21}/\rho_{11}$, $\Delta E=E_1-E_2$ and $\Delta \Gamma=\Gamma_1-\Gamma_2$. The notation $``\rm {h.c.}"$ and $``\rm{c.c.}"$ refer to Hermitian and complex conjugate, respectively. Depending on whether the eigenvalues $\lambda_m$ are pure real, we can identify three dynamical regimes associated with different dynamical behaviors of the density matrix: 

(i) When energy eigenvalues $\{\lambda_m\}$ are complex, we expect both $\Delta E\neq 0$ and $\Delta \Gamma \neq 0$. This leads to the evolution of density matrix $\hat{\rho}(t)$ exhibiting damped oscillations, due to the coexistence of a periodic function \(\Theta(t)\) and an exponential decaying factor $e^{-\Delta\Gamma t}$. 

(ii) When energy eigenvalues $\{\lambda_m\}$ are pure real, as in the case of pseudo-Hermitian systems \cite{Mostafazadeh.02.JMP}, we have $\Delta E\neq 0$ while $\Delta \Gamma=0$. In this case, Eq. \eqref{closed_rho} reduces to
\begin{equation}
\begin{aligned}
\hat{\rho}(t)~\simeq~ |\varphi_{1}\rangle\langle\varphi_{1}|  +\Theta(t).
\end{aligned}
\end{equation}
This form clearly shows that the evolution of the density matrix $\hat{\rho}(t)$ is periodic in time, with a period $T$ given by 
\begin{equation}
    T~=~\frac{2\pi}{|\Delta E|}.
    \label{closed_T}
\end{equation}

(iii) When energy eigenvalues $\{\lambda_m\}$ become pure imaginary, we have $\Delta E=0$ and $\Delta \Gamma \neq 0$, which can occur in non-Hermitian systems with an unbroken APT-symmetry. In this case, Eq. \eqref{closed_rho} simplifies to
\begin{equation}\label{eq:decay}
\begin{gathered}
    \hat{\rho}(t)~\simeq~\Theta^{\prime}  e^{-\Delta\Gamma t}.
\end{gathered}
\end{equation}
Here, the prefactor is defined as $\Theta^{\prime}=\left(R_1|\varphi_{1}\rangle\langle\varphi_{2}|+\mathrm{\text{h.c.}}\right)  - \left[\left(R_1\langle\varphi_{2}|\varphi_{1}\rangle+\mathrm{c.c.}\right)+1\right]|\varphi_{1}\rangle\langle\varphi_{1}|$. Eq. (\ref{eq:decay}) indicates that the evolution of $\hat{\rho}(t)$ now follows an exponential decay trend, with a relaxation time scale $\tau_c$ proportional to the inverse of $\Delta \Gamma$,
\begin{equation}
    \tau_c~\propto~\Delta \Gamma^{-1}.
    \label{closed_tau}
\end{equation}
We remark that Eqs. (\ref{closed_T}) and (\ref{closed_tau}) are fully determined by the spectrum of the system Hamiltonian.

We expect that information-theoretic quantities, whose dynamics are induced by the evolution of the system density matrix, may inherit the aforementioned dynamical behaviors. We would like to check whether Eqs. (\ref{closed_T}) and (\ref{closed_tau}) underlie the observed similar dynamical behaviors shared by different information-theoretic quantities. We specifically consider two information-theoretic quantities as protypes: the trace distance $D(\hat{\rho}_1(t),\hat{\rho}_2(t))$ \cite{nielsen.2001.book} that quantifies the information flow \cite{kawabata.2017.prl,P.Xue.2019.prl} and the concurrence $C(\hat{\rho}(t))$  \cite{wootters.1998.prl} that measures the entanglement \cite{Nori.2022.prr}. They are defined as (time-dependence is suppressed for simplicity)
\begin{equation}
D(\hat{\rho}_1,\hat{\rho}_2)~=~\frac12\mathrm{Tr}|\hat{\rho}_1-\hat{\rho}_2|,
\label{trace distance}
\end{equation}
and
\begin{equation}
    C(\hat{\rho})~=~\max\{0,\sqrt{r_{1}}-\sqrt{r_{2}}-\sqrt{r_{3}}-\sqrt{r_{4}}\},
\end{equation}
respectively. Here, $|\hat{A}|\equiv\sqrt{\hat{A}^\dagger\hat{A}}$, $r_j (j=1,2,3,4)$ are the eigenvalues of the matrix $R=\hat{\rho}(\sigma_y\otimes\sigma_y)\hat{\rho}^{\ast}(\sigma_y\otimes\sigma_y)$ arranged in ascending order with $\sigma_y$ denoting the Pauli matrix and $\hat{\rho}^{\ast}$ marking the complex conjugate of $\hat{\rho}$. In the following sections, we will simulate the evolution dynamics of both the trace distance and the concurrence in non-Hermitian two-qubit systems exhibiting either PT symmetry or APT symmetry to access the applicability of Eqs. (\ref{closed_T}) and (\ref{closed_tau}).    

\begin{figure}[thb!]
 \centering
\includegraphics[width=1\columnwidth]{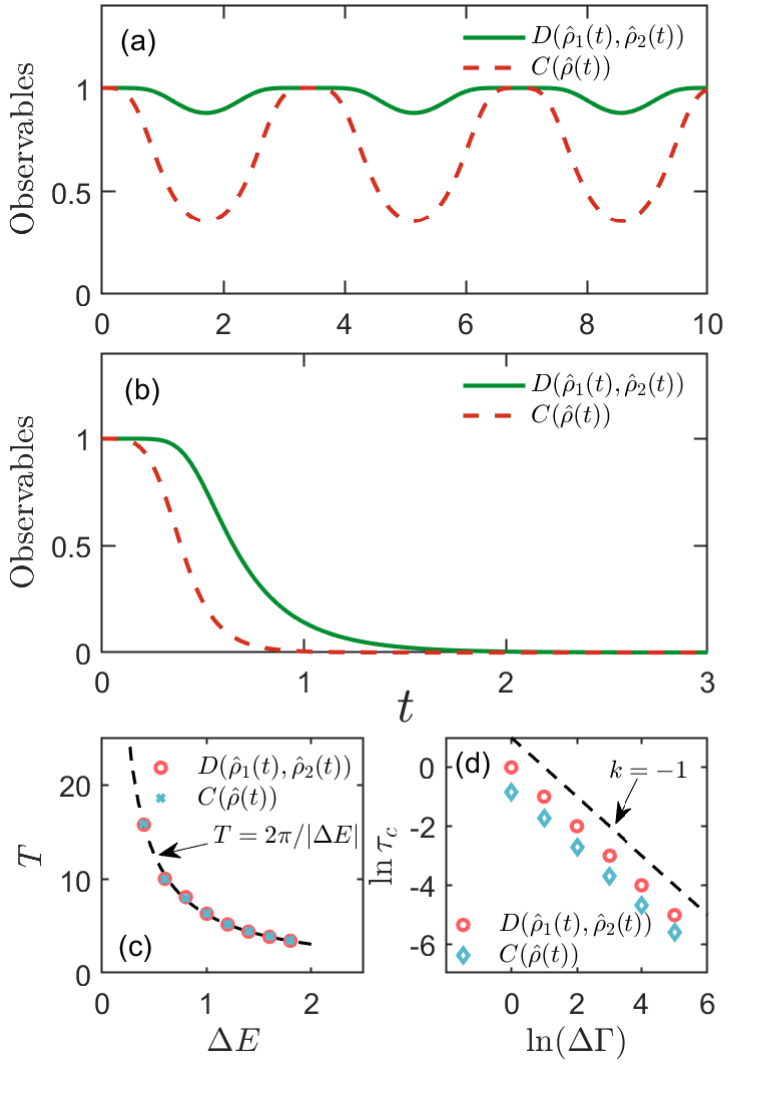} 
\caption{
Dynamics of the concurrence (red dashed line) and the trace distance (green solid line) for a non-Hermitian two-qubit system [cf. Eq. (\ref{H_PT})] in (a) the PT-unbroken phase with $a = 0.4$, and (b) the PT-broken phase with $a = 2$. In (c) and (d), we further benchmark extracted oscillation period $T$ (symbols) from (a) and relaxation times $\tau_c$ (symbols) from (b) with varying $a$ against Eq. (\ref{closed_T}) and Eq. (\ref{closed_tau}) (black dashed lines), respectively. $k=-1$ in (d) marks the slope of the dashed line. 
}
\protect\label{fig1:PT}
\end{figure}
\subsection{Non-Hermitian system with PT symmetry}\label{ssec:11}
As the first example, we examine an experimental feasible non-Hermitian two-qubit system exhibiting PT symmetry \cite{akram.2023.sr}, 
\begin{equation}
\hat{H}=\hat{H}_{1,PT}\otimes\hat{I}+\hat{I}\otimes\hat{H}_{2,PT}.
\label{H_PT}
\end{equation}
Here, $\hat{H}_{j,PT}=\hat{\sigma}_{j,x}+ia\hat{\sigma}_{j,z}$ $ (j=1,2)$ satisfying $[\hat{H}_{j,PT},\mathcal{P}\mathcal{T}]=0$ with $\mathcal{P}=\sigma_x$ and $\mathcal{T}$ representing the usual time reversal operation. We note that this single qubit Hamiltonian has been experimentally realized \cite{Fang.21.CP,Nori.2022.prr}. The operators $\hat{\sigma}_{j,x}$ and $\hat{\sigma}_{j,z}$ are Pauli matrices, while $\hat{I}$ denotes a $2\times 2$ identity matrix. We introduce ``$a$" as a tuning parameter that controls the transition between the PT-unbroken and PT-broken phases.

The eigenvalues of $\hat{H}$ can be readily obtained as
\begin{equation}
    \{\lambda_{i}\}~=~\{2\sqrt{1-a^2},0,0,-2\sqrt{1-a^2}\}.
\end{equation}
It is evident that the energy eigenvalues are real when $0<a<1$, indicating that the non-Hermitian system is in the PT-unbroken regime \cite{Bender.1998.prl}. In this scenario, we have $|\Delta E|=2\sqrt{1-a^2}$ and $\Delta \Gamma=0$ in Eqs. (\ref{closed_T}) and (\ref{closed_tau}), respectively. We thus expect purely oscillatory behavior of the density matrix. In contrast, when $a>1$, the energy eigenvalues become pure imaginary, indicating that the non-Hermitian system enters the PT-broken phase. In this phase, we have $\Delta E=0$ and $\Delta \Gamma=2\sqrt{a^2-1}$ in Eqs. (\ref{closed_T}) and (\ref{closed_tau}). We then expect that the evolution of the density operator will undergo an exponential decay process.

We now numerically check whether the dynamical behaviors of the density matrix manifest in the evolution of the trace distance and concurrence. We choose $\hat{\rho}_1(0)=|\uparrow\uparrow \rangle  \langle\uparrow\uparrow|$ and $\hat{\rho}_2(0)=|\downarrow\downarrow \rangle\langle\downarrow\downarrow|$ to simulate the evolution of the trace distance with $|\downarrow\rangle$ ($|\uparrow\rangle$) marking the ground (excited) state of $\hat{\sigma}_z$. As for the dynamics of the concurrence, we take the Bell state $|\Psi \rangle=\frac{1}{\sqrt{2}}(|\uparrow\downarrow\rangle+|\downarrow\uparrow\rangle)$ as the initial state. We choose this set of initial conditions to better illustrate the decay processes of both the trace distance and the concurrence, as both quantities start from 1 at the initial time under these conditions. We have numerically verified that our conclusions are independent of specific initial conditions. Unless otherwise stated, this set of initial conditions will be used throughout the study.

We present a set of numerical results in Fig. \ref{fig1:PT}. In Fig. \ref{fig1:PT} (a) and (b), we show dynamical evolution of the trace distance $D(\hat{\rho}_1(t),\hat{\rho}_2(t))$ and the concurrence $C(\hat{\rho}(t))$ in the PT-unbroken and PT-broken phases, respectively. One clearly observes that both the trace distance and the concurrence follow the expected dynamical trends of the density matrix in the respective regimes. 
From the dynamical curves shown in Fig. \ref{fig1:PT} (a) and (b), we can extract numerical results for the oscillation period $T$ and relaxation time $\tau_c$ for both quantities, and further benchmark the extracted data against the analytical expressions Eqs. (\ref{closed_T}) and (\ref{closed_tau}) in Fig. \ref{fig1:PT} (c) and (d), respectively. From the comparisons, we find that the extracted oscillation period in the PT-unbroken phase agrees well with the analytical expression $T=\frac{2\pi}{|\Delta E|}=\frac{\pi}{\sqrt{1-a^2}}$, and the extracted relaxation times in the PT-broken phase are indeed inversely proportional to $\Delta \Gamma=2\sqrt{a^2-1}$, as expected from Eq. (\ref{closed_tau}). Therefore, we demonstrate that the dynamical fingerprints extracted from the system density matrix indeed manifest themselves in the dynamics of information-theoretic quantities, leading to similar dynamical behaviors observed in non-Hermitian systems exhibiting PT symmetry.

\subsection{Non-Hermitian system with APT symmetry}
\begin{figure}[thb!]
 \centering
\includegraphics[width=1\columnwidth]{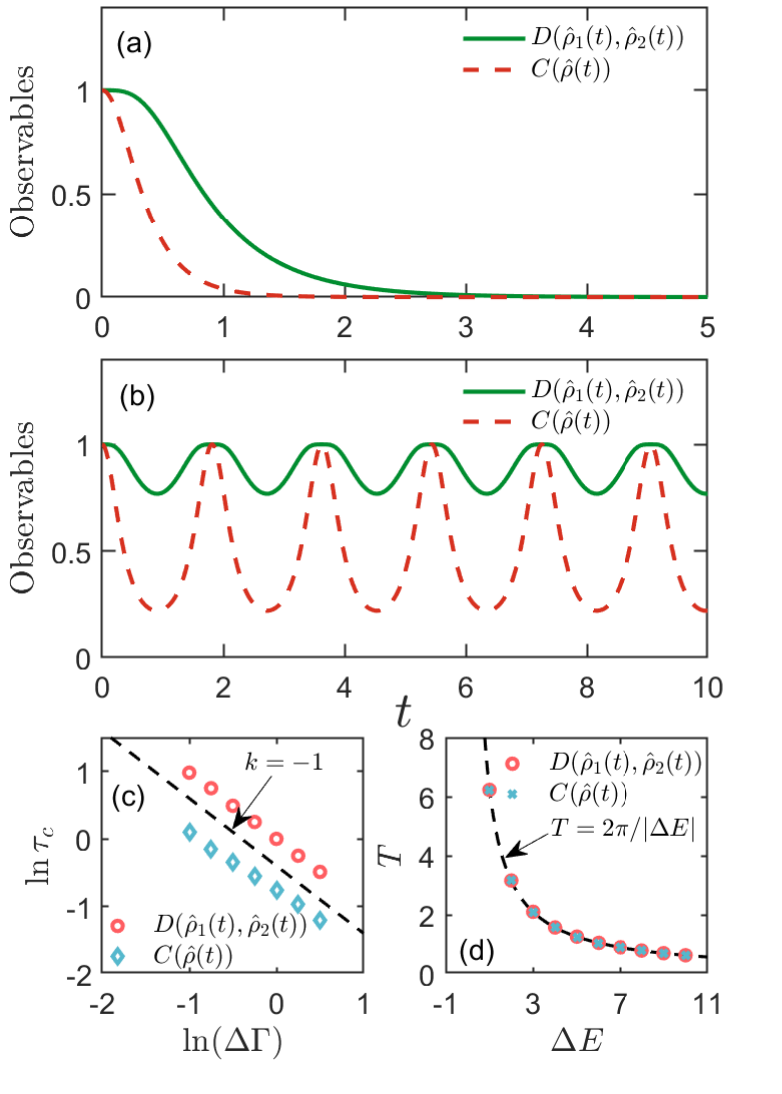} 
\caption{
Dynamics of the concurrence (red dashed line) and the trace distance (green solid line) for a non-Hermitian two-qubit system [cf. Eq. (\ref{H_APT})] in (a) the APT-unbroken phase with $b = 0.4$, and (b) the APT-broken phase with $b = 2$. In (c) and (d), we further benchmark extracted relaxation time $\tau_c$ (symbols) from (a) and oscillation periods $T$ (symbols) from (b) with varying $b$ against Eq. (\ref{closed_tau}) and Eq. (\ref{closed_T}) (black dashed lines), respectively. $k=-1$ in (c) marks the slope of the dashed line.
}
\protect\label{fig2:APT}
\end{figure}
We now turn to another important class of non-Hermitian systems that exhibit APT symmetry to assess the applicability of the dynamical fingerprints. Specifically, we consider the following non-Hermitian two-qubit system which has been experimentally realized \cite{Wen.20.NPJQI,Nori.2022.prr},
\begin{equation}
\hat{H}=\hat{H}_{1,APT}\otimes\hat{I}+\hat{I}\otimes\hat{H}_{2,APT}.
\label{H_APT}
\end{equation}
Here, $\hat{H}_{j,APT}=i\hat{\sigma}_{j,x}+b\hat{\sigma}_{j,z}$ is the Hamiltonian for $j$th qubit $(j=1,2)$ satisfying $\{\hat{H}_{j,APT},\mathcal{PT}\}=0$, with the joint operator $\mathcal{PT}$ defined above. We introduce the parameter ``$b$" to control the transition between the APT-unbroken and APT-broken phases. We emphasize that the absence of direct qubit-qubit coupling in both Eqs. (\ref{H_PT}) and (\ref{H_APT}) is intentional, based on two crucial considerations: (i) This simplified configuration significantly reduces experimental implementation complexity while maintaining the essential physics as experimentally demonstrated \cite{Wen.20.NPJQI,Nori.2022.prr}. (ii) Such design allows the composite systems to directly inherit the exact PT and APT symmetries from their constituent building blocks, thereby enabling well-defined and controllable symmetry properties in more complex non-Hermitian multi-qubit systems. This strategy eliminates the need for engineering specific qubit interactions to preserve these symmetries, which would otherwise introduce additional experimental challenges.

The eigenvalues of $\hat{H}$ can be obtained as
\begin{equation}
   \{ \lambda_{i} \}=\{2\sqrt{b^2-1},0,0,-2\sqrt{b^2-1}\}.
\end{equation}
From the above equation, we deduce that the energy eigenvalues are pure imaginary in the APT-unbroken phase with $0<b<1$. We thus obtain $|\Delta E|=0$ and $\Delta \Gamma=2 \sqrt{1-b^2}$ in Eqs. \eqref{closed_T} and \eqref{closed_tau}, respectively. Consequently, we expect pure exponential decay, as indicated by Eq. \eqref{closed_tau}. Conversely, when $b>1$, the energy eigenvalues become pure real, signaling that the non-Hermitian system enters the APT-broken phase. In this phase, we find $|\Delta E|=2\sqrt{b^2-1}$ and $\Delta \Gamma=0$ in Eqs. \eqref{closed_T} and Eqs. \eqref{closed_tau}, respectively. Thus, we expect that the evolution of the system density matrix will follow a pure oscillation trend, as described by Eq. \eqref{closed_T}.

We aim to determine whether information-theoretic quantities still inherit the dynamical fingerprints from the density matrix in the presence of the APT symmetry. To this end, we simulate the dynamical evolution of the system as given by Eq. (\ref{H_APT}) using Eq. (\ref{closed_L}) and get results for time-dependent trace distance and concurrence. A set of dynamical results for both quantities is illustrated in Fig. \ref{fig2:APT}. From Fig. \ref{fig2:APT}
(a) and (b), we observe that both the trace distance and the concurrence still exhibit dynamical behaviors similar to those of non-Hermitian systems with PT symmetry. The only difference is that the relaxation and oscillation now occur in the symmetry-unbroken and -broken phases, respectively, as opposed to the scenario in Sec. \ref{ssec:11}.

To compare with analytical expectations given by Eqs. (\ref{closed_tau}) and (\ref{closed_T}), we extract
the numerical values of relaxation time $\tau_c$ and oscillation period $T$ for both quantities from Fig. \ref{fig2:APT}
(a) and (b), respectively. From the comparisons shown in Fig. \ref{fig2:APT} (c) and (d), we find that the extracted relaxation times in the APT-unbroken phase are indeed inversely proportional to $\Delta \Gamma=2\sqrt{1-b^2}$, as expected from Eq. (\ref{closed_tau}). The extracted oscillation period in the APT-broken phase also agrees well with the analytical expression $T=\frac{2\pi}{|\Delta E|}=\frac{\pi}{\sqrt{b^2-1}}$. 

Together with results showed in Sec. \ref{ssec:11}, we demonstrate that the dynamical fingerprints as described by Eqs. (\ref{closed_T}) and (\ref{closed_tau}) underpin the dynamical behaviors of both the trace distance and the concurrence regardless of the symmetry type. This finding highlights a form of dynamical universality in non-Hermitian systems relevant to information-theoretic quantities. Thus, we provide a unified framework to understand existing separate efforts \cite{Nori.2022.prr,akram.2023.sr,kawabata.2017.prl,P.Xue.2019.prl,DingYL.2022.pra} that focused on individual types of quantities and systems. Although we focus here on the trace distance and the concurrence, we emphasize that our conclusion extends naturally to all information-theoretic quantities whose dynamics are solely attributed to the system density matrix in closed non-Hermitian systems.

\section{Pure dephasing effect} \label{sec:3}
Although the revealed dynamical fingerprints are fascinating due to their generality and simplicity, we should always bear in mind that real-world imperfections are very often hindering the beauty uncovered under ideal conditions. This holds true for non-Hermitian systems as well: realistic non-Hermitian systems are typically prepared through trajectory selection or truncation over dissipative larger systems \cite{Ozdemir.19.NM,chenwj.2021.prl,chenwj.2022.prl,ren.2022.np,sun.2023.pra,Bergholtz.21.RMP}, implying that non-Hermitian systems should be subject to external perturbations. Since information-theoretic quantities rely on quantum features that are fragile under detrimental external perturbations, an urgent question arises about the fate of the previously identified dynamical fingerprints in open non-Hermitian systems. To provide insights, we focus on the effect of pure dephasing, which is one of leading obstacles that limits the current quantum applications.

\subsection{Quantum master equation and trivial pure dephasing effect}
Treating non-Hermitian systems as open ones, the Hamiltonian of the composite system can be generally expressed as 
\begin{equation} \hat{H}_{\rm{tot}}=\hat{H}\otimes\hat{I}_E+\hat{I}_S\otimes\hat{H}_E+\hat{H}_{I}.
\end{equation}
Here, $\hat{H}$ and $\hat{H}_E$ are the Hamiltonians of non-Hermitian systems and the environment, respectively. $\hat{H}_{I}=\sum_a\gamma_a\hat{C}_a\otimes\hat{E}_a$ describes the system-bath coupling, with $\hat{C}_a$ and $\hat{E}_a$ representing the system and environmental operators, respectively. 

Since the composite system $\hat{H}_{\rm{tot}}$ is closed, its evolution is governed by Eqs. \eqref{closed_H} and  \eqref{closed_L} as well. For simplicity, we here assume weak coupling between non-Hermitian systems and their environments. By tracing out the environmental degrees of freedom, $\hat{\rho}=\mathrm{Tr}_E\hat{\rho}_{\rm{tot}}$, and invoking the Born-Markov approximations, one can derive the following quantum master equation that governs the evolution of open non-Hermitian systems (see Appendix \ref{a:2} for derivation details)
\bea\label{open_L}
   \frac{d}{dt}\hat{\rho}(t) &=& -i\left[\hat{H}\hat{\rho}(t)-\hat{\rho}(t)\hat{H}^{\dagger}\right]+\sum_k\mathcal{D}_k[\hat{\rho}(t)]\nonumber\\
   &&-i\mathrm{Tr}[\hat{\rho}(t)(\hat{H}^{\dagger}-\hat{H})]\hat{\rho}(t).
\eea
Compared to Eq. (\ref{closed_L}), we observe that Eq. (\ref{open_L}) includes an extra dissipator $\sum_k\mathcal{D}_k[\hat{\rho}(t)]$ with $\mathcal{D}_k[\hat{\rho}(t)]\equiv \gamma_{k}\left[\hat{L}_{k}\hat{\rho}(t)\hat{L}_{k}^{\dagger}-\frac{1}{2}\{\hat{L}_{k}^{\dagger}\hat{L}_{k},\hat{\rho}(t)\}\right]$ which captures the dissipation effects induced by the environment; $\gamma_k$ marks the damping coefficient of channel $k$, $\hat{L}_{k}$ represents a Lindblad jump operator and $\{\hat{A},\hat{B}\}$ is the anti-commutator between the two operators $\hat{A},\hat{B}$. We note that similar forms of Eq. (\ref{open_L}) have appeared in prior studies \cite{sun.2023.pra,Yuan.20.PRL,Xu.24.PRL}. However, we remark that Ref. \cite{sun.2023.pra} obtained the form in a specific setting and missed the correction term such that the conservation of probability is violated as already noted by the authors of Ref. \cite{sun.2023.pra}. While Refs. \cite{Yuan.20.PRL,Xu.24.PRL} utilized the complete form of Eq. (\ref{open_L}) for specific applications, a justified derivation was lacking.  

Here we focus on pure dephasing effects and particularly distinguish the local pure dephasing which acts locally on the two qubits and the collective counterpart which treats the two qubits as a whole. In our analysis, we explicitly consider the following form of the dissipator:
\begin{equation}
\sum_k\mathcal{D}_k[\hat{\rho}(t)]=\sum_{k=1,2,3}\gamma_{k}\left(\hat{L}_{k}\hat{\rho}(t)\hat{L}_{k}^{\dagger}-\frac{1}{2}\{\hat{L}_{k}^{\dagger}\hat{L}_{k},\hat{\rho}\}\right).
\end{equation}
Here, we set $\hat{L}_{1}=\sigma_{1,z}\otimes I$ and $\hat{L}_{2}=I \otimes \sigma_{2,z}$, which act locally on the first and second qubit, respectively. $\hat{L}_{3}=\sigma_{1,z}\otimes I+I\otimes\sigma_{2,z}$ describes the collective dephasing effect. In simulations, we modify the type of pure dephasing by varying values of $\gamma_k$.

Before proceeding, it is important note that Eq. (\ref{open_L}) includes several known equations of motion as special cases.  When $\hat{H}=\hat{H}^{\dagger}$, the correction term in the second line of Eq. (\ref{open_L}) vanishes, recovering the usual quantum Lindblad master equation frequently utilized for addressing open Hermitian systems. When $\gamma_k=0$, Eq. (\ref{open_L}) reduces to Eq. (\ref{closed_L}) for closed non-Hermitian systems. When both conditions $(\hat{H}=\hat{H}^{\dagger}$ and $\gamma_k=0)$ are satisfied, we obtain the von Neumann equation for closed Hermitian systems. 

\begin{figure}[b!]
 \centering
\includegraphics[width=1\columnwidth]{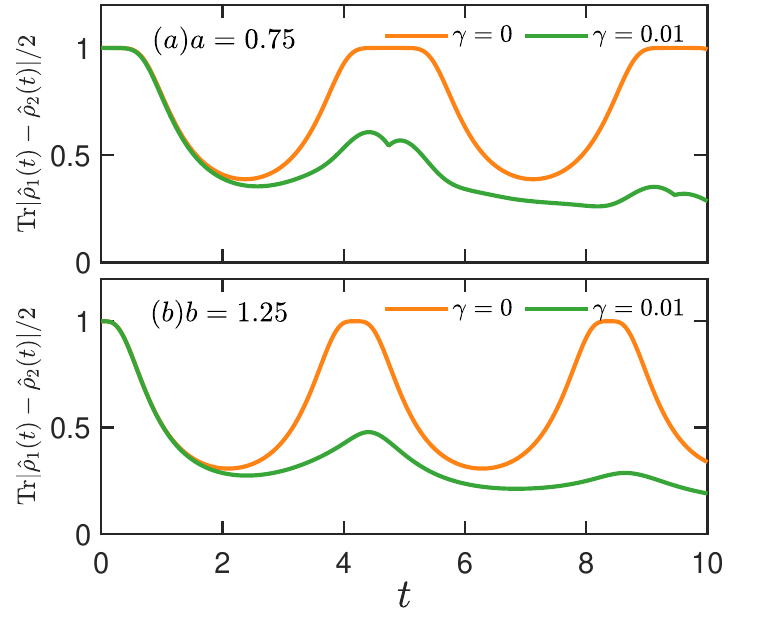} 
\caption{
Comparisons between the evolution dynamics of the trace distance in (a) the PT-unbroken regime with $a=0.75$ and (b) the APT-broken regime with $b=1.25$ with and without the collective pure dephasing. Other parameters for the orange and green solid lines are $\gamma_1=\gamma_2= \gamma_3=\gamma=0$ and $\gamma_1=\gamma_2=0, \gamma_3=\gamma = 0.01$, respectively.
}
\protect\label{fig:trace distance}
\end{figure}

The formal solution of Eq. \eqref{open_L} can be expressed as \cite{campaioli.2024.prx}
\begin{equation}
\hat{\rho}(t)=\frac{\mathrm{vec}^{-1}\left(e^{\mathcal{L}_0 t}\mathrm{vec}(\hat{\rho}(0))\right)}{\mathrm{Tr}[\mathrm{vec}^{-1}\left(e^{\mathcal{L}_0 t}\mathrm{vec}(\hat{\rho}(0))\right)]}.
\label{open_H}
\end{equation}
Here, $\mathrm{vec}(\mathcal{O})$ transforms an operator $\mathcal{O}$ into a column vector, while $\mathrm{vec}^{-1}$ is its inverse operation that converts a column vector back into matrix form. We denote the Liouvillian superoperator as $  \mathcal{L}_0 \hat{\rho}(t)\equiv-i\left(\hat{H}\hat{\rho}(t)-\hat{\rho}(t)\hat{H}^{\dagger}\right)+\sum_k\mathcal{D}_k(\hat{\rho}(t))$. The term $e^{\mathcal{L}_0t}$ represents the matrix exponential of the Liouvillian superoperator, encapsulating the time evolution of the system in vectorized space. The denominator 
$\mathrm{Tr}[\mathrm{vec}^{-1}\left(e^{\mathcal{L}t}\mathrm{vec}(\rho(0))\right)]$ ensures proper normalization of the density matrix and serves as the correction term $-i\mathrm{Tr}[\hat{\rho}(t)(\hat{H}^{\dagger}-\hat{H})]\hat{\rho}(t)$ in Eq. \eqref{open_L}. Eq. (\ref{open_H}) indicates that the evolution of the system's density matrix can be fully determined by solving the following eigenproblem of $\mathcal{L}_0$ 
\begin{equation}
    \mathcal{L}_0 \mathrm{vec}(\hat{\rho}_j)~=~\mu_j \mathrm{vec}(\hat{\rho}_j).
\end{equation}
Here, $\{\hat{\rho}_j\}$  and $\{\mu_j\}=\{\mathcal{E}_j+i\eta_j\}$ represent the corresponding eigenmatrices and eigenvalues of the superoperator $\mathcal{L}_0$.
We sort $\hat{\rho}_j$ in descending order with respect to the real parts $\{ \mathcal{E}_{j} \}$, i.e., $\mathcal{E}_{1} \geq \mathcal{E}_{2} \geq \mathcal{E}_{3} \geq \cdots$. Indeed, the steady state of Eq. \eqref{open_L} corresponds exactly to $\hat{\rho}_1$ despite the presence of the nonlinear correction term in Eq. \eqref{open_L} (see Appendix \ref{a:3} for details). 

Since the eigenvalues of $\mathcal{L}_0$ are generally complex, we expect the oscillatory evolution trend of the density matrix in closed non-Hermitian systems to be significantly modified by the presence of pure dephasing. In Fig. \ref{fig:trace distance}, we compare the evolution dynamics of the trace distance in both the PT-unbroken and APT-broken regimes, with and without pure dephasing. As expected, perfect oscillations in closed non-Hermitian systems become damped oscillations in open non-Hermitian systems subjected to pure dephasing. We have numerically verified that the oscillation dynamics of the concurrence experiences similar changes under the action of pure dephasing. Hence, in the dynamical regime of non-Hermitian systems where information-theoretic quantities can depict perfect oscillations in closed settings, the pure dephasing just behaves similarly to its Hermitian counterpart by suppressing oscillations.

\subsection{Nontrivial pure dephasing effect on relaxation}
\begin{figure}[b!]
 \centering
\includegraphics[width=1\columnwidth]{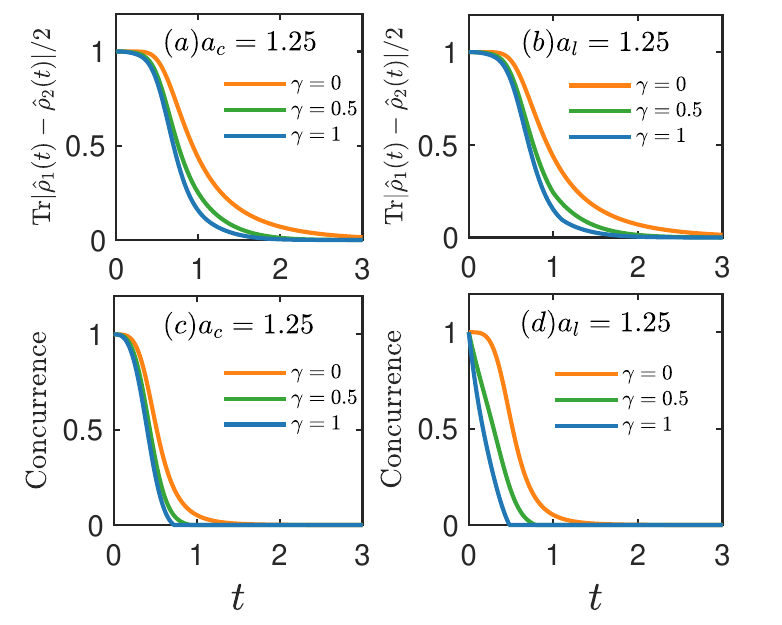} 
\caption{Evolution of trace distance [(a)(b)] and concurrence [(c)(d)] in the PT-broken regime with $a=1.25$ under varying strength of pure dephasing. The subscripts ``$c$" and ``$l$" of ``$a$" denote the collective and local pure dephasing with $(\gamma_1=\gamma_2=0, \gamma_3=\gamma)$ and $(\gamma_1=\gamma_2=\gamma, \gamma_3=0)$, respectively. 
}
\protect\label{fig4:openPT}
\end{figure}

We now turn to the dynamical regime of non-Hermitian systems where information-theoretic quantities already depict decaying behaviors and examine the impact of pure dephasing. We find that the pure dephasing can induce unexpected anomalous behaviors that are absent in Hermitian counterparts. Since the relaxation time is of primary interest in this dynamical regime, we first analyze that of Eq. \eqref{open_L}. Referring to Eq. \eqref{open_H}, we can express the Liouvillian gap \cite{Mori.2023.prl} in terms of just eigenvalues of $\mathcal{L}_0$, which reads $\Delta\mathcal{E}\equiv \mathcal{E}_1-\mathcal{E}_2$. We note that $\mathcal{E}_1$ vanishes when considering Hermitian systems. Indeed, we show that the defined Liouvillian gap determines the asymptotic behavior of the density matrix near the steady state (see details in Appendix \ref{a:3})
\begin{equation}
    \hat{\rho}(t)~\simeq~\hat{\rho}_1 + R \cdot  e^{-\Delta \mathcal{E} t} e^{-i \Delta \eta t}.
    \label{tau_open}
\end{equation}
Here, $R$ is a parameter matrix related to the initial state of the system, and $\Delta \eta=\eta_1-\eta_2$. In terms of the Liouvillian gap, we can define the relaxation time governed by Eq. (\ref{open_L}) as
\begin{equation}\label{eq:tauo}
    \tau_o=\frac{1}{\Delta \mathcal{E}}.
\end{equation}

We first analyze non-Hermitian systems in the PT-broken regime, using the system Hamiltonian provided in Eq. (\ref{H_PT}). We summarize a set of typical results in Fig. \ref{fig4:openPT}. In Fig. \ref{fig4:openPT} (a) and (b), we present the evolution dynamics of the trace distance under varying strength of collective and local pure dephasing, respectively. Fig. \ref{fig4:openPT} (c) and (d) illustrate the evolution dynamics of the concurrence under varying strengths of collective and local pure dephasing, respectively. For both information-theoretic quantities, we clearly observe that the pure dephasing just accelerates the relaxation dynamics in the PT-broken regime, regardless of its type. The stronger the pure dephasing, the faster the relaxation process occurs. This is a somewhat expected consequence of pure dephasing \cite{breuer.2002.book}. As the coupling strength increases, the detrimental influence of the environment becomes more significant, resulting in a faster loss of quantum features.

We then address the impact of pure dephasing on non-Hermitian systems described by Eq. (\ref{H_APT}) in the APT-unbroken regime. 
\begin{figure}[thb!]
 \centering
\includegraphics[width=1\columnwidth]{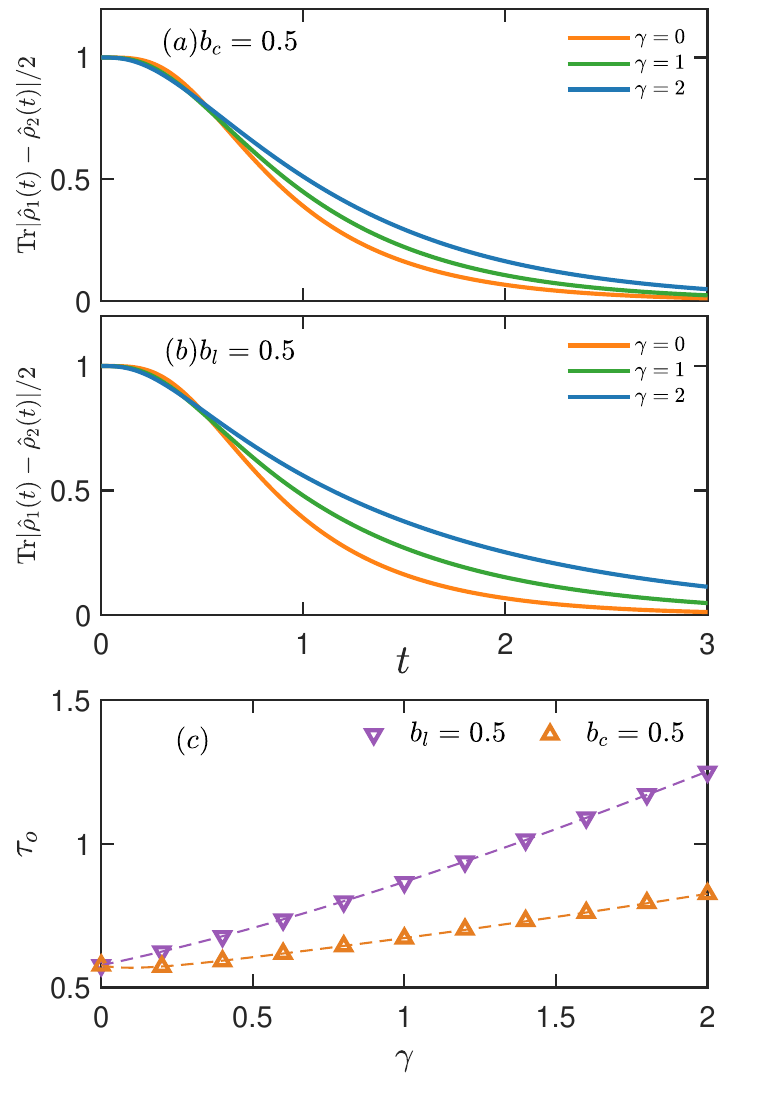} 
\caption{The anomalous slow-down effect of pure dephasing: 
The evolution dynamics of trace distance in the APT-unbroken regime of Eq. (\ref{H_APT}) with $b=0.5$ under varying strength of (a) collective pure dephasing with $(\gamma_1=\gamma_2=0, \gamma_3=\gamma)$ and (b) local pure dephasing with $(\gamma_1=\gamma_2=\gamma, \gamma_3=0)$. (c) The extracted relaxation time of the system $\tau_o$ with varying $\gamma$ from (a) (orange symbols) and (b) (purple symbols). The dashed lines are the theoretical predictions from Eq. (\ref{eq:tauo}).
}
\protect\label{fig4:openAPT_trace distance}
\end{figure}
We will show that an anomalous slow-down effect of pure dephasing emerges, with no counterparts in both Hermitian systems and PT-symmetric non-Hermitian systems. To elaborate on the results in depth, we first focus on the dynamical response of the trace distance to pure dephasing in the APT-unbroken regime. A prototypical set of dynamical results for the trace distance is shown in Fig. \ref{fig4:openAPT_trace distance}. Interestingly, we observe from Fig. \ref{fig4:openAPT_trace distance} (a) and (b) that the presence of pure dephasing, whether it is collective or local, can suppress the relaxation of the trace distance compared to the scenario without pure dephasing. The extracted values for the relaxation time $\tau_o$ from the dynamical trends shown in (a) and (b) are presented in Fig. \ref{fig4:openAPT_trace distance} (c). We also compare the extracted data with the theoretical predictions using Eq. (\ref{eq:tauo}) and find perfect agreements between them. Thus, we identify an intriguing slow-down effect of pure dephasing.

To provide analytical insights into this anomalous effect of pure dephasing in APT-symmetric systems, we consider a strong dephasing limit of $\gamma \gg 1$. In this limit, we can neglect the quantum coherence of the density matrix and focus on the dynamical behavior of its diagonal elements. Applying this limit to the quantum master equation Eq. (\ref{open_L}) for the APT-symmetric non-Hermitian systems [cf. Eq. (\ref{H_APT})], we find that the diagonal elements $\rho_{nn}=\langle n|\hat{\rho}|n\rangle$ ($n=1,2,3,4$) in the basis of $\{|1\rangle=|\uparrow\uparrow\rangle,|2\rangle=|\uparrow\downarrow\rangle,|3\rangle=|\downarrow\uparrow\rangle,|4\rangle=|\downarrow\downarrow\rangle\}$ satisfy the following equations of motion (see details in Appendix \ref{a:4}) 
\begin{equation}
    \dot{\rho}_{nn}(t)~=~0.
    \label{APT_theta0_m}
\end{equation}
Here, $\dot{\rho}_{nn}\equiv d\rho_{nn}/dt$. This simple form directly implies that the dynamics of the density matrix is completely frozen in the strong dephasing limit. We remark that this suppression phenomenon is distinct from quantum Zeno effect typically observed in open Hermitian systems, as Eq. (\ref{APT_theta0_m}) arises from the interplay between pure dephasing and APT symmetry. In Appendix \ref{a:5}, we further contrast numerical results for complex eigenvalues of Liouvillian superoperators of non-Hermitian and Hermitian single-qubit systems to highlight such a dynamical distinction. Consequently, the relaxation of the trace distance that builds upon the density matrix should be completely ceased in that limit. Considering the exponential relaxation in closed systems, one would naturally expect a slow-down phenomenon to occur when increasing the strength of pure dephasing. We note that the simple form of Eq. (\ref{APT_theta0_m}) disappears immediately when considering non-Hermitian systems with PT symmetry. 

\begin{figure}[thb!]
 \centering
\includegraphics[width=1\columnwidth]{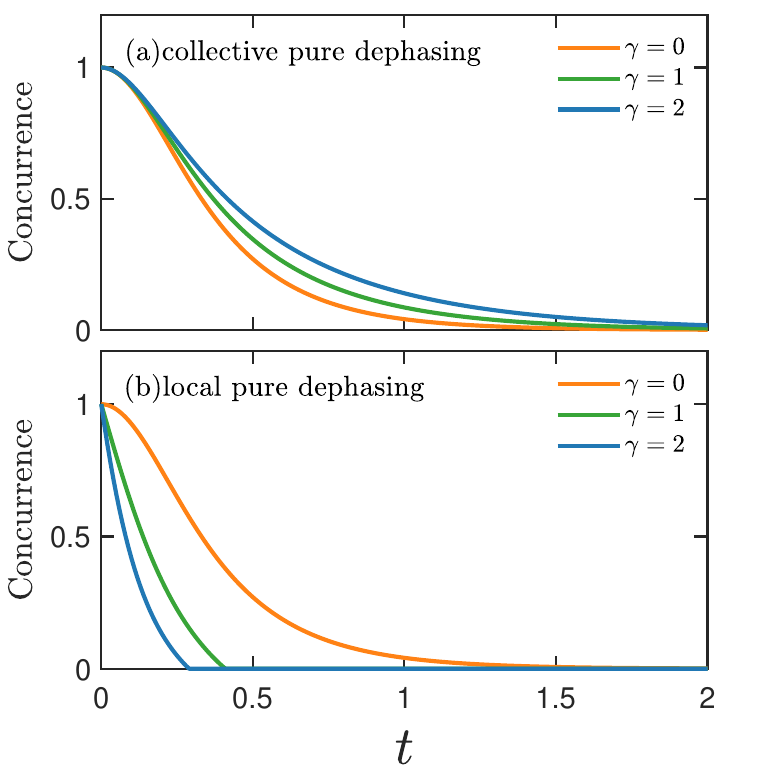} 
\caption{The evolution of the concurrence over time in APT symmetric system [cf. Eq. (\ref{H_APT})] under (a) collect pure dephasing and (b) local pure dephasing with varying strength. Other parameters are the same with Fig. \ref{fig4:openAPT_trace distance}.}
\protect\label{fig5:openAPToncurrence}
\end{figure}

We then examine the scenario of the concurrence, with a set of results shown in Fig. \ref{fig5:openAPToncurrence}. In the presence of collective dephasing, we observe from Fig. \ref{fig5:openAPToncurrence} (a) that a slow-down effect, similar to that of the trace distance, still exists. However, unlike the trace distance, the concurrence is sensitive to the type of pure dephasing. In Fig. \ref{fig5:openAPToncurrence} (b), we show that local dephasing just accelerates rather than suppresses the decay of the concurrence, in stark contrast to the trace distance. We attribute this difference to the fact that local dephasing can easily destroy the global entanglement quantified by concurrence \cite{aolita.2015.iop}, leading to the absence of a slow-down effect.  

Therefore, by selecting an appropriate type of pure dephasing, one can observe a slow-down effect on the inherent relaxation of information-theoretic quantities induced by pure dephasing in APT-symmetric non-Hermitian systems, which is completely absent in non-Hermitian systems with PT symmetry by varying the strength of pure dephasing. Our results thus provide strong evidence for a general passive protection strategy for APT-symmetric non-Hermitian systems by harnessing pure dephasing.

\subsection{Extent of dephasing induced slow-down phenomenon}
Given the potential applications of the dephasing-induced slow-down phenomenon in quantum information protection, a crucial question arises regarding its persistence in more complex APT-symmetric non-Hermitian systems beyond the two-qubit case studied numerically. We affirmatively address this question through the following theoretical construction. We first note that the most general APT-symmetric Hamiltonian of a single qubit takes the form
\begin{equation}\label{eq:apt_g}
    \hat{H}^g_{APT}=
\left[\begin{array}{cc}
    b+i\theta & i\kappa-s\\
    i\kappa+s & -b+i\theta\end{array}\right].
\end{equation}
The single qubit model utilized in Eq. (\ref{H_APT}) corresponds to the special choice $\theta=0=s$ and $\kappa=1$. With this single-qubit Hamiltonian in Eq. (\ref{eq:apt_g}), we can construct a APT-symmetric multi-qubit model through the tensor-product extension 
\begin{equation}\label{eq:apt_tot}
    \hat{H}=\sum_{j=1}^N\hat{H}_{j,APT}^g.
\end{equation}
Here, each $\hat{H}_{j,APT}^g$ takes the form in Eq. (\ref{eq:apt_g}) and $N\ge 1$ can be an arbitrary integer. We emphasize that the proposed multi-qubit model constitutes a natural extension of the experimentally realized two-qubit system described by Eq. (\ref{H_APT}). This direct generalization ensures that the implementation of such multi-qubit systems remains well within reach of current experimental capabilities, particularly given recent advances in the control of non-Hermitian quantum systems. 

In the limit of strong dephasing applied either locally or collectively, we can just focus on the dynamics of diagonal elements of system density matrix in the local basis constructed from $\{|\uparrow\rangle,|\downarrow\rangle\}^{\otimes N}$. Remarkably, we still find a complete dynamical freezing as dictated by the following vanishing equations of motion (see details in Appendix \ref{a:4}) 
\begin{equation}
   \dot{\rho}_{nn}(t)=0.
\end{equation}
Here, $n\in(1,2,\cdots,N)$. This result demonstrates that the dynamical freezing phenomenon persists in an APT-symmetric multi-qubit system given by Eq. (\ref{eq:apt_tot}) under strong dephasing, generalizing our earlier two-qubit findings. Importantly, this implies that all information-theoretic quantities derived solely from the system density matrix will necessarily inherit a dephasing induced protection mechanism in this APT-symmetric multi-qubit architecture. Moreover, the scalability of this architecture suggests that the observed dynamical phenomena can be systematically investigated across different system sizes using existing quantum simulation platforms.

\section{Conclusion}\label{sec:4}
In this study, we presented a comprehensive analysis of dynamical non-Hermitian systems in both closed and open settings, focusing primarily on the dynamical behaviors of information-theoretic quantities. In closed non-Hermitian systems, we identified symmetry-independent dynamical fingerprints that underpin similar dynamical behaviors observed in distinct information-theoretic quantities, with the trace distance and the concurrence as specific examples. We thus unified disparate dynamical results from previous studies \cite{Nori.2022.prr,akram.2023.sr,kawabata.2017.prl,P.Xue.2019.prl,DingYL.2022.pra} on various non-Hermitian systems and information-theoretic quantities, deepening our understanding of the commonalities in the dynamical behaviors of non-Hermitian systems across different symmetries. 

Turning to open non-Hermitian systems experiencing pure dephasing, we demonstrated a surprising slow-down effect of pure dephasing in APT-symmetric non-Hermitian systems, in that pure dephasing can suppress the inherent relaxation of information-theoretic quantities. We emphasize that the dephasing-induced slow-down phenomenon reported here differs fundamentally from similar effects observed in previous studies \cite{brody2012mixed,Gardas.2016.pra} in several key aspects: First, while Refs. \cite{brody2012mixed,Gardas.2016.pra} investigated systems with PT symmetry, we demonstrate a protection effect induced by dephasing in non-Hermitian systems with unbroken APT symmetry. Second, regarding the underlying physics, Ref. \cite{brody2012mixed} showed how white noise removes phase transitions by inducing a finite order parameter in the phase-unbroken regime whereas Ref. \cite{Gardas.2016.pra} focused on protecting quantum coherence (or equivalently, slowing dephasing itself), without establishing whether such protection extends to other physical quantities. On the contrary, we revealed a protection mechanism arising from the interplay between APT symmetry and dephasing that systematically slows the evolution dynamics of density matrix, thereby ensuring protection for all quantities built solely upon the density matrix, as we explicitly demonstrated through both trace distance and concurrence analyses. The ability of APT-symmetric non-Hermitian systems to benefit from the presence of pure dephasing could lead to new information protection strategies that can promote quantum information applications of non-Hermitian devices. Since our adopted qubit models are well within the current experimental capabilities, we expect direct experimental verifications of our findings in the near future.

\section*{Acknowledgments}
J.L. acknowledges support from the National Natural Science Foundation of China (Grant No. 12205179), the Shanghai Pujiang Program (Grant No. 22PJ1403900) and start-up funding of Shanghai University.

\appendix

\renewcommand{\theequation}{A\arabic{equation}}
\setcounter{equation}{0}  
\section{Density matrix of closed non-Hermitian systems in the asymptotic limit} \label{a:1}
For non-Hermitian systems with Hamiltonian $\hat{H}_{}$,  the initial density matrix is defined as follows:
\begin{equation}
\hat{\rho}(0)=\sum_{mn}\rho_{mn}\left|\varphi_m\right\rangle\left\langle\varphi_n\right|,
\end{equation}
where $\left|\varphi_n\right\rangle (\left\langle\chi_{n}\right|)$ is a right (left) eigenstate of $\hat{H}$ with eigenvalue $\lambda_{n}$, and the elements of density matrix is $\begin{aligned}\rho_{mn}&=\frac{\langle\chi_m|\hat{\rho}_0|\chi_n\rangle}{\langle\chi_m|\varphi_m\rangle\langle\varphi_n|\chi_n\rangle}\end{aligned}$ with $\langle\chi_m|\varphi_n\rangle=\delta_{mn}$. The density matrix at time $t$ can be obtained as $\hat{\rho}(t)=\hat{U}(t)\hat{\rho}(0)\hat{U}^\dagger(t)/\mathrm{Tr}\left[ \hat{U}(t)\hat{\rho}(0)\hat{U}^\dagger(t)\right]=e^{-i\hat{H}t}\hat{\rho}(0)e^{i\hat{H}^{\dagger}t}/\mathrm{Tr}\left[ e^{-i\hat{H}t}\hat{\rho}(0)e^{i\hat{H}^{\dagger}t}\right]$ \cite{brody2012mixed,cao.2023.prr,du.2022.pra,P.Xue.2019.prl,Nori.2022.prr,Yuan.20.PRL,Mao.24.CPL}. In terms of the eigenspectrum of $\hat{H}$, we can express the density matrix at finite time as
\begin{equation}\label{eq:a2}
    \hat{\rho}\left(t\right)=\frac{\sum_{mn}\rho_{mn}e^{-i\omega_{mn}t}\left|\varphi_{m}\right\rangle\left\langle\varphi_{n}\right|}{\mathrm{Tr}[\sum_{mn}\rho_{mn}e^{-i\omega_{mn}t}\left|\varphi_{m}\right\rangle\left\langle\varphi_{n}\right|]}.
\end{equation}
Here, we have denoted $\omega_{mn}=\lambda_{m}-\lambda_{n}$. For later convenience, we sort the eigenvalues $\lambda_m=E_m+i\Gamma_m$ in descending order of $\Gamma_m$, such that $\Gamma_1 > \Gamma_2 >\cdots$. 

To proceed, we transfer Eq. (\ref{eq:a2}) into
\begin{equation}
    \hat{\rho}\left(t\right)=\frac{\sum_{mn}\frac{\rho_{mn}}{\rho_{11}}e^{-i(E_{m}-E_{n})t}e^{\left(\Gamma_{m}+\Gamma_{n}-2\Gamma_1\right)t}\left|\varphi_{m}\right\rangle\left\langle\varphi_{n}\right|}{\mathrm{Tr}\left[\sum_{mn}\frac{\rho_{mn}}{\rho_{11}}e^{-i(E_{m}-E_{n})t}e^{\left(\Gamma_{m}+\Gamma_{n}-2\Gamma_1\right)t}\left|\varphi_{m}\right\rangle\left\langle\varphi_{n}\right| \right]}
\end{equation}
by multiplying both the numerator and denominator with a common factor $e^{-2\Gamma_{1}t}/\rho_{11}$. We then consider the long time limit where only the steady state and the term containing the smallest decaying rate $\Delta \Gamma=\Gamma_1-\Gamma_2$ survive
\bea
\hat{\rho}\left(t\right) &\simeq& \frac{\left|\varphi_{1}\right\rangle\left\langle\varphi_{1}\right|+\left[R_1 e^{-i\Delta E t}\left|\varphi_{1}\right\rangle\left\langle\varphi_{2}\right|+\mathrm{\text{h.c.}}\right]e^{-\Delta\Gamma\cdot t}}{\mathrm{Tr}\left[\left|\varphi_{1}\right\rangle\left\langle\varphi_{1}\right|+\left[R_1 e^{-i\Delta E t}\left|\varphi_{1}\right\rangle\left\langle\varphi_{2}\right|+\mathrm{\text{h.c.}}\right]e^{-\Delta\Gamma\cdot t}\right]} \nonumber\\
&=& \frac{\left|\varphi_1\right\rangle\left\langle\varphi_1\right|+\left[R_1 e^{-i\Delta E t}\left|\varphi_1\right\rangle\left\langle\varphi_2\right|+\mathrm{h.c.}\right]e^{-\Delta\Gamma\cdot t}}{1+\left[R_1 e^{-i\Delta E t}\left\langle\varphi_2|\varphi_1\right\rangle+\mathrm{c}.\mathrm{c}.\right]e^{-\Delta\Gamma\cdot t}}.
\eea
Here, $\Delta E=E_1-E_2$, $R_1=\rho_{12}/\rho_{11}$, and the notations ``h.c." and ``c.c." refer to Hermitian and complex conjugate, respectively. Invoking the approximation $\frac{1}{1+x}\approx1-x$ valid under $|x|<<1$, we ultimately obtain
\begin{equation}
\begin{gathered}
    \hat{\rho}(t)\simeq |\varphi_{1}\rangle\langle\varphi_{1}|  +\Theta(t) e^{-\Delta\Gamma\cdot t},
\end{gathered}
\end{equation}
where we have defined 
\bea
    \Theta(t) &=& \{\left[R_1e^{-i\Delta E t}|\varphi_{1}\rangle\langle\varphi_{2}|+\mathrm{\text{h.c.}}\right] \nonumber \\
    &&- \left[R_1e^{-i\Delta E t}\langle\varphi_{2}|\varphi_{1}\rangle+\mathrm{c.c.}\right]|\varphi_{1}\rangle\langle\varphi_{1}|\}.
\eea

\renewcommand{\theequation}{B\arabic{equation}}
\setcounter{equation}{0}  
\section{Quantum master equation for open non-Hermitian systems}
\label{a:2}
Consider the composite Hamiltonian $\hat{H}_{\mathrm{tot}}$ describing a non-Hermitian system $\hat{H}$ interacting with an environment $\hat{H}_{E}$ 
\begin{equation}
\hat{H}_{\mathrm{tot}}=\hat{H}\otimes\hat{I}_{E}+\hat{I}_{S}\otimes\hat{H}_{E}+\hat{H}_{I}.
\end{equation}
Here, $\hat{H}_E$ denotes the Hamitonian of environment which is assumed to be Hermitian, the interaction Hamiltonian $\hat{H}_{I}=\sum_{a}\gamma_a\hat{C}_a\otimes\hat{E}_a$ with $\hat{C}_a$ and $\hat{E}_a$ system and environmental operators, respectively. In the interaction picture, the density matrix of the composite system satisfies
\begin{equation}
    \frac{d}{dt}\hat{\rho}_{\mathrm{tot}}^{I}(t)=-i[\hat{V}_{I}\left(t\right)\hat{\rho}_{\mathrm{tot}}^{I}(t)-\hat{\rho}_{\mathrm{tot}}^{I}(t)\hat{V}_{I}^{\dagger}(t)].
    \label{eq:rhoSB_I}
\end{equation}
Here, we have denoted $\hat{\rho}_{\mathrm{tot}}^{I}(t)=e^{i\hat{H}_{\mathrm{eff}}t}\hat{\rho}_{\mathrm{tot}}(t)e^{-i\hat{H}_{\mathrm{eff}}^{\dagger}t}$ with  $\hat{H}_{\mathrm{eff}} \equiv\hat{H}\otimes\hat{I}_{E}+\hat{I}_{S}\otimes\hat{H}_{E}$, $\hat{V}_{I}(t)=e^{i\hat{H}_{\mathrm{eff}}t}\hat{H}_{I}e^{-i\hat{H}_{\mathrm{eff}}^{\dagger}t}$ is the interaction Hamiltonian in the interaction picture.

The formal solution of Eq. \eqref{eq:rhoSB_I} reads
\begin{equation}
\begin{aligned}
    \hat{\rho}_{\mathrm{tot}}^{I}(t)=&\hat{\rho}_{\mathrm{tot}}^{I}(0)\\
    &-i\int_{0}^{t}ds\left[\hat{V}_{I}\left(s\right)\hat{\rho}_{\mathrm{tot}}^{I}\left(s\right)-\hat{\rho}_{\mathrm{tot}}^{I}\left(s\right)\hat{V}_{I}^{\dagger}\left(s\right)\right].
\end{aligned}
    \label{eq:rhoSB_I2}
\end{equation}
Following the standard procedures of open quantum system theory \cite{breuer.2002.book} that lead to perturbative quantum master equation, one can get \cite{du.2022.pra}
\begin{equation}
    \begin{aligned}
        \frac{d}{dt}\hat{\rho}^{I}(t) = & \, \mathrm{Tr}_E \Bigg[ \int_{0}^{t} ds \, \hat{V}_{I}(t-s) \left(\hat{\rho}^{I}(t) \otimes \hat{\rho}_{E}\right) \hat{V}_{I}^{\dagger}(t)  \\
& - \int_{0}^{t} ds \, \hat{V}_{I}(t) \hat{V}_{I}(t-s) \left(\hat{\rho}^{I}(t) \otimes \hat{\rho}_{E}\right) \Bigg] + \text{h.c.}. 
    \end{aligned}
    \label{interect density}
\end{equation}
Here, $\hat{\rho}_{E}$ is the initial state for the environment.

Recall we consider the right-state time evolution \cite{Mao.24.CPL}. In terms of the right eigenstates of $\hat{H}$, the interaction Hamiltonian in the interaction picture can be expressed as
\begin{equation}
  \hat{V}_I\left(t\right)=\sum_a\sum_\omega e^{-i\omega t}\hat{A}_a(\omega)\otimes\hat{E}_a(t).
\end{equation}
Here, we have defined $\hat{E}_a(t)=e^{i\hat{H}_{E}t}\hat{E}_{a}e^{-i\hat{H}_{E}t}$ and
\begin{equation} \hat{A}_a(\omega)=\gamma_a\sum_m|\varphi_{m}\rangle\left\langle\varphi_{m}\right|\hat{C}_a|\varphi_{m+\omega}\rangle\left\langle\varphi_{m+\omega}\right|.
\end{equation}
The action of this operator on the system induces transition from state $|\varphi_{m+\omega}\rangle$ to state $|\varphi_{m}\rangle$. It satisfies the property $\hat{A}_a(\omega)=\hat{A}^{\dagger}_a(-\omega)$.

Substituting the interaction Hamiltonian into Eq. \eqref{interect density}, we get
\begin{equation}
\begin{aligned}
\frac{d}{dt}\rho^{I}(t)=&\sum_{a,b}\sum_{\omega,\omega_1}e^{i(\omega_1-\omega)t}\hat{A}_b(\omega)\hat{\rho}^I(t)\hat{A}_a^\dagger(\omega_1)  \\
& \times \int_0^tds e^{i\omega s}\mathrm{Tr}_E\left(\hat{E}_a^\dagger(t)\hat{E}_b(t-s)\hat{\rho}_E\right)  \\
&-\sum_{a,b}\sum_{\omega,\omega_1}e^{i(\omega_1-\omega)t}\hat{A}^{\dagger}_a(\omega)\hat{A}_b(\omega)\hat{\rho}^I(t)  \\
&\times \int_0^tdse^{i\omega s}\mathrm{Tr}_E\left(\hat{E}_a^\dagger(t)\hat{E}_b(t-s)\hat{\rho}_E\right)+\text{h.c.}.
\end{aligned}
\end{equation}
We define the reservoir correlation function $\Gamma_{ab}\left(\omega\right)=\int_{0}^{t}dse^{i\omega s}\mathrm{Tr}_{E}\left(\hat{E}_{a}^{\dagger}(t)\hat{E}_{b}(t-s)\rho_{E}\right)$ and rearrange the above equation as
\begin{equation}
\begin{aligned}
\frac{d}{dt}\hat{\rho}^{I}(t)=&\sum_{a,b}\sum_{\omega,\omega_{1}}e^{i(\omega_{1}-\omega)t}\Gamma_{ab}(\omega)\left[\hat{A}_{b}(\omega)\hat{\rho}^I(t)A_a^{\mathrm{\dagger}}(\omega_{1}) \right.\\
&\left. -\hat{A}_a(-\omega_1)\hat{A}_b(\omega)\hat{\rho}^I(t)\right]+\text{h.c.}.
\end{aligned}
\end{equation}

We further employ the rotating wave approximation which amounts to neglecting the transitions $\omega\neq\omega_1$, yielding
\begin{equation}
\begin{aligned}
\frac{d}{dt}\hat{\rho}^{I}(t) =& \sum_{a,b}\sum_{\omega}\left[\Gamma_{ab}\left(\omega\right)(\hat{A}_{b}(\omega)\hat{\rho}^{I}(t)\hat{A}_{a}^{\dagger}(\omega)  \right. \\
 &\left. -\hat{A}_a(-\omega)\hat{A}_b(\omega)\hat{\rho}^{I}(t))+\text{h.c.}\right].
\end{aligned}
\label{eq:interact rho}
\end{equation}

Leaving behind the normalization factor for a moment, the equation of motion for the density matrix in the Schr\"odinger picture can be obtained by converting that in the interaction picture according to the following form
\begin{equation}
    \frac{d}{dt}\hat{\rho}(t) =e^{-i\hat{H}t}\left(\frac{d}{dt}\hat{\rho}^{I}(t)\right) e^{i\hat{H}^\dagger t}-i \Big[\hat{H}\hat{\rho}(t)-\hat{\rho}(t)\hat{H}^\dagger\Big].
    \label{eq:schro rho}
\end{equation}
By substituting Eq. \eqref{eq:interact rho} into \eqref{eq:schro rho}, we can obtain
\begin{equation}
\begin{aligned}
     \frac{d}{dt}\hat{\rho}(t) =&\sum_{a,b}\sum_{\omega}\left[\Gamma_{ab}\left(\omega\right)(\hat{A}_{b}(\omega)\hat{\rho}(t)\hat{A}_{a}^{\dagger}(\omega) \right.\\
&\left. -\hat{A}^{\dagger}_a(\omega)\hat{A}_b(\omega)\hat{\rho}(t))+\text{h.c.}\right]\\
&-i \Big[\hat{H}\hat{\rho}(t)-\hat{\rho}(t)\hat{H}^\dagger\Big].
\end{aligned} 
\end{equation}
To simplify the above form, we decompose the operators $\Gamma_{ab}$ into Hermitian and non-Hermitian parts $\Gamma_{ab}=\frac{1}{2}\gamma_{ab}+i S_{ab}$ such that the above equation becomes
\begin{equation}
\begin{aligned}
     \frac{d}{dt}\hat{\rho}(t) =&\sum_{a,b}\sum_{\omega} \frac{1}{2}\gamma_{ab}\left[\hat{A}_{b}(\omega)\hat{\rho}(t)\hat{A}_{a}^{\dagger}(\omega)+\hat{A}_{a}(\omega)\hat{\rho}(t)\hat{A}_{b}^{\dagger}(\omega) \right.\\
& \left.- \hat{A}_{a}^{\dagger}(\omega)\hat{A}_{b}(\omega)\hat{\rho}(t)-\hat{\rho}\hat{A}^{\dagger}_b(\omega)\hat{A}_a(\omega) \right]\\
& -i\Big[(\hat{H}+\hat{H}^{\dagger}_{\mathrm{Ls}})\hat{\rho}(t)-\hat{\rho}(t)(H^{\dagger}+H^{\dagger}_{\mathrm{Ls}})\Big].
\end{aligned}
\end{equation}
Here, we have denoted the Lamb shift Hamiltonian  $\hat{H}^{\dagger}_{\mathrm{Ls}}=\sum_{a,b}\sum_{\omega}S_{ab}(\omega)\hat{A}_{a}^{\dagger}(\omega)\hat{A}_{b}(\omega)$. At weak couplings, we can neglect the effects of the Lamb shift. We can further cast the above equation into a quantum master equation of a Lindblad form
\begin{equation}
\begin{aligned}
\frac d{dt}\hat{\rho}(t)=&-i\Big[\hat{H}\hat{\rho}(t)-\hat{\rho}(t)\hat{H}^\dagger\Big]+\sum_k\mathcal{D}_{k}\Big[\hat{\rho}(t)\Big].\\
\end{aligned}
\label{eq17}
\end{equation}
Here, we have defined a dissipator $ \mathcal{D}_{k}\Big[\hat{\rho}(t)\Big]\equiv \gamma_k[L_{k}\hat{\rho}(t) L_{k}^{\dagger}-\frac{1}{2}L_{k}^{\dagger}L_{k}\hat{\rho}(t)-\frac{1}{2}\hat{\rho}(t) L_{k}^{\dagger}L_{k}]$ which captures the dissipation effects induced by the environment. By performing the trace operation over both sides of Eq. \eqref{eq17}, one can find that
\begin{equation}
    \frac d{dt}\left[\mathrm{Tr}\hat{\rho}(t)\right]=i\mathrm{Tr}\left[\hat{\rho}(t) (\hat{H}^\dagger-\hat{H})\right] \neq 0.
\end{equation}
To ensure the conservation of probability, we can add a correction term to the master equation that compensates the above contribution \cite{brody2012mixed}, thereby arriving at the following complete form of quantum master equation used for investigating open non-Hermitian systems in the main text
\begin{equation}
\begin{aligned}
    \frac d{dt}\hat{\rho}(t)=&-i\left[\hat{H}\hat{\rho}(t)-\hat{\rho}(t) \hat{H}^\dagger\right]+\sum_k\mathcal{D}_{k}\Big[\hat{\rho}(t)\Big]  \\
    &-i\mathrm{Tr}\left[\hat{\rho}(t) (\hat{H}^{\dagger}-\hat{H})\right]\hat{\rho}(t).
    \label{eq:open_master_eq}
\end{aligned}
\end{equation}

\renewcommand{\theequation}{C\arabic{equation}}
\setcounter{equation}{0}  
\section{Density matrix of open non-Hermitian systems }\label{a:3}
By defining the superoperator $\mathcal{L}_{0}(\hat{\rho}(t))=-i\left[ \hat{H}\hat{\rho}(t)-\hat{\rho}(t) \hat{H}^\dagger\right]+\sum_k\mathcal{D}_{k}\Big[\hat{\rho}(t)\Big]$, the formal solution of Eq. \eqref{eq:open_master_eq} can be expressed as
\begin{equation}
    \hat{\rho}(t)=\frac{\mathrm{vec}^{-1}\left(e^{\mathcal{L}_0t}\mathrm{vec}(\hat{\rho}(0))\right)}{\mathrm{Tr}[\mathrm{vec}^{-1}\left(e^{\mathcal{L}_0t}\mathrm{vec}(\hat{\rho}(0))\right)]}.
\label{eq18}
\end{equation}
Here, vec$(\mathcal{O})$ transforms an operator $\mathcal{O}$ into a column vector, while vec$^{-1}$ is its inverse operation that converts a column vector back into matrix form. The denominator of Eq. (\ref{eq18}) plays a role of the correction term. The form of Eq. (\ref{eq18}) indicates that one can utilize the eigenspectrum of the superoperator $\mathcal{L}_{0}$ to analyze the density matrix,
\begin{equation}
    \mathcal{L}_0 \mathrm{vec}(\hat{\rho}_j)=\mu_j \mathrm{vec}(\hat{\rho}_j).
\end{equation}
Here, $\{\mu_j=\mathcal{E}_j+i\eta_j\}$ and $\{\hat{\rho}_j\}$ are eigenvalues and eigenmatrices of the superoperator $\mathcal{L}_{0}$, respectively. We sort $\{\hat{\rho}_j\}$ in descending order according to the real part of the eigenvalues, i.e., $\mathcal{E}_{1} \geq \mathcal{E}_{2} \geq \mathcal{E}_{3} \geq \cdots$. The normalized density matrix reads
\begin{equation}
   \hat{\rho}(t)=\frac{\sum_{j}e^{\mu_j t} C_j \hat{\rho}_j}{\mathrm{Tr}\left(\sum_{j}e^{\mu_j t} C_j \hat{\rho}_j \right)}.
\end{equation}
Extracting $C_1 e^{\mu_1 t}$ from both the numerator and denominator, $\hat{\rho}(t)$ can be given by
\begin{equation}
    \begin{aligned}
\hat{\rho}(t)=&\frac{\sum_{j\geq2}C_{1}e^{\mu_{1}t}\left(\hat{\rho}_{1}+\frac{C_{j}}{C_{1}}e^{-\left(\mu_{1}-\mu_{j}\right)t}\hat{\rho}_{j} \right)}{\mathrm{Tr}\left[\sum_{j\geq2}C_{1}e^{\mu_{1}t}\left(\hat{\rho}_{1}+\frac{C_{j}}{C_{1}}e^{-\left(\mu_{1}-\mu_{j}\right)t}\hat{\rho}_{j} \right) \right]}\\
=&\frac{\sum_{j\geq2}\left( \hat{\rho}_{1}+\frac{C_{j}}{C_{1}}e^{-\left(\mu_{1}-\mu_{j}\right)t}\hat{\rho}_{j}\right)}{\sum_{j\geq2}\left(1+\frac{C_{j}}{C_{1}}e^{-\left(\mu_{1}-\mu_{j}\right)t}\right)}
\end{aligned}
\end{equation}

In the long time limit, only the contribution of $\hat{\rho}_1$ survives, and thus the steady-state $\hat{\rho}_{ss}$ is given by
\begin{equation}
    \hat{\rho}_{ss}~=~  \hat{\rho}_1.
\end{equation}
Therefore, the steady state of open non-Hermitian system is determined by the eigenmatrix of the superoperator $ \mathcal{L}_0$ with the largest real part. 

We then consider asymptotic relaxation dynamics of the density matrix towards the steady state. To this end, we consider states that deviate from the steady state with a nonnormalized form reading
\begin{equation}
    \hat{\rho}(t)\propto  \hat{\rho}_{1}+\sum_{j\geq2}\frac{C_{j}}{C_{1}}e^{-\left(\mu_{1}-\mu_{j}\right)t}\hat{\rho}_{j}.
\end{equation}
Here, we have utilized the symbol ``$\propto$" to emphasize the nonnormalized nature of the state. 
From this form, one directly find that the asymptotic relaxation dynamics of state that is close to the steady state is described by
\begin{equation}
    \hat{\rho}(t)\propto\hat{\rho}_1 + R e^{-\Delta \mathcal{E} t} e^{-i \Delta \eta t}.
\end{equation}
Here, $\Delta \mathcal{E}=\mathcal{E}_1-\mathcal{E}_2$ corresponds to the Liouvillian gap, $\Delta \eta=\eta_1-\eta_2$ and $R=C_2\hat{\rho}_2/C_1$ is determined by the initial state by noting that $\hat{\rho}(0)\propto\sum_{j\geq1}C_j \hat{\rho}_j$.

\renewcommand{\theequation}{D\arabic{equation}}
\setcounter{equation}{0}  
\section{Strong dephasing limit of APT symmetric systems} \label{a:4}
\subsection{Two-qubit model}
We first consider the quantum master equation Eq. (\ref{eq:open_master_eq}) describing a two-qubit model with APT symmetry in the strong dephasing limit of $\gamma_k\gg1$. In this limit, we can neglect the off-diagonal elements of $\hat{\rho}(t)$ and focus on just diagonal ones. We consider the local basis $\{|1\rangle=|\uparrow\uparrow\rangle,|2\rangle=|\uparrow\downarrow\rangle,|3\rangle=|\downarrow\uparrow\rangle,|4\rangle=|\downarrow\downarrow\rangle\}$ and explicitly write down the equations of motion for the diagonal elements $\rho_{nn}(t)=\langle n|\hat{\rho}(t)|n\rangle$ ($n=1,2,3,4$),
\begin{equation}
    \begin{aligned}
   \frac{d}{dt}\rho_{nn}(t)=&-i [H_{nn}\rho_{nn}(t)-\rho_{nn}(t)H^{\dagger}_{nn}]+\langle n|\sum_k\mathcal{D}_k[\hat{\rho}(t)]|n\rangle\\
&-i\sum_{m}\left(\rho_{mm}(t) H^{\dagger}_{mm}-\rho_{mm}(t)H_{mm}\right)\rho_{nn}(t).
\label{rho_dig}
    \end{aligned}
\end{equation}


Here we limit to pure dephasing such that $\hat{L}_1=\sigma_{1,z}\otimes I$, $\hat{L}_2=I \otimes \sigma_{2,z}$ and  $\hat{L}_3=\sigma_{1,z}\otimes I +I \otimes \sigma_{2,z}$ with $I$ the $2\times2$ identity. 
We note that the diagonal elements of $\hat{L}_1$, $\hat{L}_2$, $\hat{L}_3$ in the local basis are real, $\hat{L}_{1,nn}=\hat{L}^{\dagger}_{1,nn}$, $\hat{L}_{2,nn}=\hat{L}^{\dagger}_{2,nn}$, $\hat{L}_{3,nn}=\hat{L}^{\dagger}_{3,nn}$, we find
\begin{equation}
\begin{aligned}
    \left[\frac{d}{dt}\rho_{nn}(t)\right]_{(2)}\equiv\langle n|\sum_k\mathcal{D}_k[\hat{\rho}(t)]|n\rangle=0,
\end{aligned}\label{rho_2}
\end{equation}
where $\left[\frac{d}{dt}\rho_{nn}(t)\right]_{(n)}$ represents the $n$-th term on the right side of Eq. \eqref{rho_dig}. For the APT symmetric Hamiltonian considered in Eq. (\ref{H_APT}), we also have real diagonal elements, $\hat{H}_{nn}=  \hat{H}_{nn}^{\dagger}$, leading to
\begin{equation}
      \left[\frac{d}{dt}\rho_{nn}(t)\right]_{(1)}=\left[\frac{d}{dt}\rho_{nn}(t)\right]_{(3)}=0.
\end{equation}
Putting together, we finally get
\begin{equation}
    \frac{d}{dt}{\rho}_{nn}(t)~=~0.
    \label{APT_theta0}
\end{equation}
This equation indicates the dynamics of the density matrix of APT symmetric systems is completely frozen in the strong dephasing limit, providing an explanation of why a slow down effect of pure dephasing should occur in this kind of non-Hermitian systems. We remark that $\hat{H}_{nn}\neq\hat{H}_{nn}^{\dagger}$ in general for non-Hermitian system with PT symmetry as described by Eq. (\ref{H_PT}), hence the simple equation of motion Eq. (\ref{APT_theta0}) is absent.

\subsection{Generalized multi-qubit model}
We now consider an APT-symmetric multi-qubit systems $\hat{H}=\sum_{j=1}^N\hat{H}_{j,APT}^{g}$ defined in Eq. (\ref{eq:apt_tot}) of the main text. In the strong dephasing limit, the diagonal elements of the system density matrix in the local basis is still governed by Eq. (\ref{rho_dig}) with Hamiltonian replaced by the generalized one. As we consider pure dephasing and the Hamiltonian still respects the APT symmetry, the part of equation of motion $\left[\frac{d}{dt}\rho_{nn}(t)\right]_{(2)}\equiv\langle n|\sum_k\mathcal{D}_k[\hat{\rho}(t)]|n\rangle$ can be easily showed to be vanishing, just as Eq. (\ref{rho_2}). Moreover, we note that the diagonal elements of single-qubit Hamiltonian reads $\langle \uparrow|H_{j,APT}^{g}|\uparrow\rangle=b+i\theta$ and $\langle \downarrow|H_{j,APT}^{g}|\downarrow\rangle=-b+i\theta$. Thus, the remaining components of the equation of motion $d\rho_{nn}/dt$ read
\bea
\left[\frac{d}{dt}\rho_{nn}(t)\right]_{(1)} &\equiv& -i[H_{nn}\rho_{nn}(t)-\rho_{nn}(t)H_{nn}^{\dagger}]\nonumber\\
  &=& 2N\theta \rho_{nn}(t),\nonumber\\
\left[\frac{d}{dt}\rho_{nn}(t)\right]_{(3)} &\equiv & -i\sum_{m}\left(\rho_{mm}(t)H_{mm}^{\dagger}-\rho_{mm}(t)H_{mm}\right)\rho_{nn}(t)\nonumber\\
  &=& -2N\theta \rho_{nn}(t).
\label{APT_theta}
\eea
Altogether, we still find that 
\begin{equation}
\frac{d}{dt}\rho_{nn}(t)~=~0
\end{equation}
for the defined APT-symmetric multi-qubit system. This result indicates that the freezing of dynamics in the strong dephasing limit is a general feature of considered APT-symmetric qubit systems. Hence, we expect that the dephasing induced slow-down of relaxation of information-theoretic quantities numerically observed in a two-qubit model persists in generalized APT-symmetric multi-qubit systems.
\begin{figure}[thb!]
 \centering
\includegraphics[width=1\columnwidth]{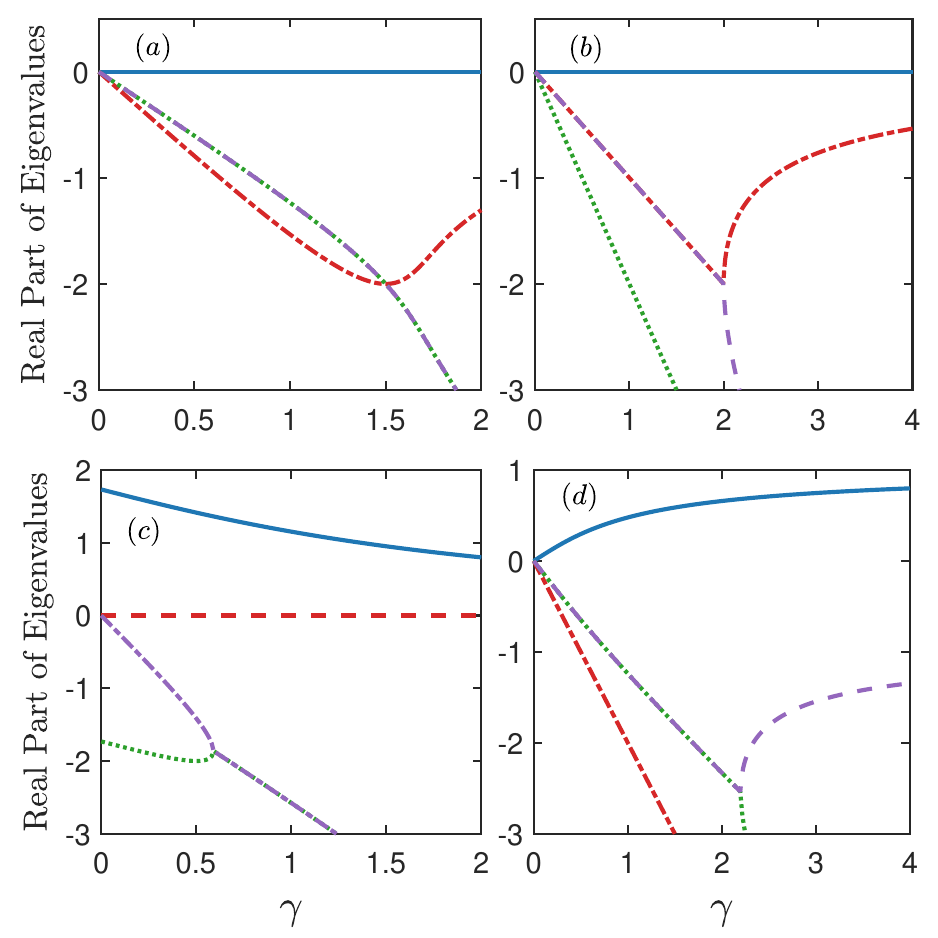} 
\caption{Real parts of all eigenvalues of Liouvillian as a function of pure dephasing strength $\gamma$. We contrast four single-qubit scenarios with Hermitian Hamiltonian (a) $H=\sigma_x+b\sigma_z$ and (b) $H=\sigma_x$, as well as non-Hermitian Hamiltonian (c) $H=i\sigma_x+b\sigma_z$ and (d) $H=\sigma_x+ia\sigma_z$ that respect an APT and PT symmetry, respectively. To align with results showed in the main text, we take $a=b=0.5$.
}
\label{fig:real}
\end{figure}

\renewcommand{\theequation}{E\arabic{equation}}
\setcounter{equation}{0}  
\section{Spectrum of Liouvillian superoperator} \label{a:5}
To complement dynamical results for information-theoretic quantities showed in the main text, here we present numerical results for the eigenvalue spectrum of the Liouvillian superoperator $\mathcal{L}_0$ defined below Eq. (\ref{open_H}) of the main text in the presence of pure dephasing. Particularly, we also contrast eigenvalues of Liouvillian between Hermitian and non-Hermitian systems which provide unambiguous evidence differentiating conventional quantum Zeno effects in Hermitian systems from the dephasing induced slow-down phenomenon we report in APT-symmetric non-Hermitian systems.

To provide insights while maintaining clarity and simplicity, we focus on the single-qubit scenario where the Liouvillian spectrum consists of exactly four eigenvalues. We perform a systematic comparison between our adopted non-Hermitian systems and two representative Hermitian systems: (i) $H=\sigma_x+b\sigma_z$, chosen for its structural similarity to our APT-symmetric Hamiltonian in Eq. (\ref{H_APT}), and (ii) $H=\sigma_x$, representing the Hermitian limit $(a\to 0)$ of our PT-symmetric model in Eq. (\ref{H_PT}). Fig. \ref{fig:real} displays the real parts of the corresponding Liouvillian eigenvalues, which provide direct access to two crucial dynamical features: (i) the Liouvillian gap (defined as the difference between real parts of two eigenvalues having the largest real parts), and (ii) its dependence on dephasing strength $\gamma$. This analysis reveals fundamental differences in how Hermitian and non-Hermitian systems respond to decoherence.

From the figure, we infer that the Liouvillian gap of both open Hermitian systems [Fig. \ref{fig:real} (a) and (b)] and open APT-symmetric non-Hermitian system [Fig. \ref{fig:real} (c)] tends to vanishing in the strong dephasing limit, implying that the dynamics of system density matrix is frozen. We note that the behavior in the former is conventionally referred to as quantum Zeno effect. However, we emphasize that the non-Hermitian systems demonstrate qualitatively distinct features that preclude a simple identification of the observed dephasing induced slow-down phenomenon with the Zeno effect, as evidenced by:

(i) The Liouvillian gap exhibits markedly different behaviors in the weak dephasing regime for open Hermitian and APT-symmetric non-Hermitian systems. Most notably, the APT-symmetric non-Hermitian system [Fig. \ref{fig:real} (c)] displays a strictly monotonic decrease of the Liouvillian gap with increasing dephasing strength--a trend that aligns perfectly with the dynamical results shown in Fig. \ref{fig4:openAPT_trace distance}. This stands in sharp contrast to the non-monotonic dependence observed for both Hermitian systems [Fig. \ref{fig:real} (a) and (b)], where the Liouvillian gap initially increases before exhibiting decreasing features at intermediate dephasing strengths. These fundamental differences in gap evolution directly reflect the distinct underlying mechanisms governing the dynamics in Hermitian versus APT-symmetric non-Hermitian systems.  

(ii) The systems shown in Fig. \ref{fig:real} (b) and (d) are related through continuous variation of the non-Hermiticity parameter $a$, yet they exhibit strikingly different dynamical behaviors. While the Hermitian system ($a=0$) demonstrates conventional quantum Zeno behavior under strong dephasing [Fig. \ref{fig:real} (b)], the PT-symmetric non-Hermitian system maintains a finite Liouvillian gap in the strong dephasing limit [Fig. \ref{fig:real} (d)], as our extensive numerical calculations confirmed. This persistent gap indicates the absence of complete dynamical freezing, in sharp contrast to the Hermitian case. This contrast provides compelling evidence that the quantum Zeno effect does not generically extend to non-Hermitian systems. Moreover, the marked difference between these continuously connected systems highlights how non-Hermiticity can qualitatively alter a system's response to environmental interactions.

Our analysis leads to the conclusion that the dephasing-induced slow-down phenomenon in APT-symmetric systems arises from a mechanism fundamentally distinct from the quantum Zeno effect. This novel dynamical behavior stems from the unique interplay between APT non-Hermitian symmetry and pure dephasing.

%

\end{document}